\documentclass{appolb}
\usepackage{graphicx}
\begin{document}
\title{KONDO EFFECT IN THE TRANSPORT THROUGH A QUANTUM DOT:
       EXTENDED NONCROSSING APPROXIMATION APPROACH
\thanks{Presented at the Strongly Correlated Electron Systems 
Conference, Crak\'ow 2002.}%
}
\author{D. Gerace, E. Pavarini, and L. C. Andreani
\address{INFM-Dipartimento di Fisica ``A. Volta'', 
         Universit\`a di Pavia, Via Bassi 6, I-27100 Pavia, Italy}
}
\maketitle

\begin{abstract}
We calculate the conductance through a single quantum dot 
coupled to metallic leads, modeled by the spin 1/2 Anderson model.
We adopt the finite-$U$ extension of the noncrossing approximation
method.
Our results are in good agreement with exact 
numerical renormalization group results both
in the high temperature 
and in the Kondo (low temperature) regime.
Thanks to this approach, we were able to fit fairly well
recently reported 
measurements \cite{silvano} in a quantum dot device.
We show that, contrarily to what previously suggested,
the conductance of this particular device  can be understood 
within the spin 1/2 Anderson model,
in which the effects of the multilevel structure 
of the dot are neglected.
\end{abstract}
\PACS{73.23.-b, 72.15.Qm, 73.63.Kv}

\vspace{0.5cm}


The recent observation \cite{sper,vanderwiel} of the Kondo effect 
\cite{hewson} in quantum dot (QD) devices
has opened the possibility to control the Kondo effect experimentally,
and a new exciting field of research \cite{kouw}.
It was shown that -- below a Kondo temperature 
$T_K$ of the order of 100 mK -- the linear response
conductance of a QD device, G, approaches $2e^2/h$ (unitary limit) if
the number $n$ of electrons confined in the QD is odd, and 
it becomes very small if $n$ is even \cite{vanderwiel}. 
This even/odd effect seems to be a general feature of 
most \cite{sper2} QD devices.
In the present paper we
focus our attention on those devices for which Kondo anomalies
appear \textit{only} for odd $n$ \cite{silvano,sper,vanderwiel}, 
and in particular on the experimental data reported 
very recently in Ref. \cite{silvano}.

In order to describe a QD coupled to its leads, we adopt the 
spin 1/2 Anderson model \cite{teor}
\begin{eqnarray}
\label{anderson}
\!\!\!\!\!\! H\!\!&=&\!\!\!\!\!\!\!\sum_{(\mathbf{k},\sigma)\in S,D}
       \!\! \!\!\varepsilon_{\mathbf{k}}\,
        c_{{\mathbf{k}}\sigma}^{\dagger}
        c_{{\mathbf{k}}\sigma}\!+\!
\varepsilon_0\sum_{\sigma}\,d_{\sigma}^{\dagger}d_{\sigma}+ U
n_{d\uparrow} n_{d\downarrow}
\!+\!\!\!\!\!\sum_{({\mathbf{k}},\sigma)\in S,D } \!\!\!
      \left(V_{{\mathbf{k}}\sigma}
            c_{{\mathbf{k}}\sigma}^{\dagger}
            d_{\sigma}+
            h.c.\right)\! .
\end{eqnarray}

Here $c_{\mathbf{k}\sigma}^\dagger$ ($c_{\mathbf{k}\sigma}$)
creates (destroys) a conduction electron with momentum
$\mathbf{k}$, energy $\varepsilon_{\mathbf{k}}$ and spin 
$\sigma$ in one of the two leads, which we
label with S (source) and D (drain); $d_{\sigma}^\dagger$
($d_{\sigma}$) creates (destroys) an electron with
spin $\sigma$ on the QD;
$V_{{\mathbf{k}}\sigma}$ is the hybridization between QD and
conduction states (which we suppose to be $\mathbf{k}$ 
independent, $|V_{{\mathbf{k}}\sigma}|=V_{S(D)}$ for $\mathbf{k}\in S$ or
$D$), $\varepsilon_0$ is the energy of a single electron localized on
the QD and $U$ is the Coulomb interaction among electrons on the
same orbital level. In the present model the energy $\varepsilon_0$ is 
not fixed, but on the contrary 
it is tuned by a gate voltage, $V_g$,
coupled to the QD through a capacitor. 
As a first approximation we assume 
a linear relation, $\varepsilon_0=\alpha V_g+const$.
The gate voltage (and thus $\varepsilon_0$) 
controls the number of electrons confined in the QD at low temperature.
In the linear response regime ($V_{SD}\ll V_g$, where $V_{SD}$ 
is the source-drain bias) 
the conductance, G, may be written in a Landauer-like
form \cite{meir}
\begin{equation}
\label{landauer} {\rm
G}(T,V_g)=\frac{2e^2}{h}\int_{-\infty}^{+\infty}\pi\Gamma\left(-
\frac{1}{\pi}{\rm Im}\{G^R(\varepsilon+ i\eta)\}\right)
\left(-\frac{\partial{f}}{\partial{\varepsilon}}\right)\,{\rm
d}\varepsilon.
\end{equation}
\noindent
Here  we have assumed for simplicity that the couplings 
to the leads are symmetric. 
The actual QD-leads coupling strength is  
thus $\Gamma=\pi N(\varepsilon_F)V^2$, where 
$V=(V_S^2+V_D^2)^{1/2}$, and
$N(\varepsilon_F)$ is the density of states per spin at the
Fermi level \cite{teor}.
In addition, $f$ is the Fermi distribution 
function and $G^R(\varepsilon+ i\eta)$
is the retarded local Green function, i.e. the Fourier transform of
the time-dependent function 
$G^R(t)=-{\rm i}\,\theta(t)\langle|\{d(t),d^{\dagger}(0)\}|\rangle$.
In order to calculate the local density of states (DOS), that is
$\rho=-$Im$\{G^R(\varepsilon+ i\eta)\}/{\pi}$, we adopt the noncrossing 
approximation method in its finite-$U$ extension (UNCA) 
\cite{pruschke}.
This method allows us to explore the empty orbital,
($n\sim 0$, $-\varepsilon_0<-\Gamma$), 
the Kondo ($n\sim 1$, $\Gamma<-\varepsilon_0<U-\Gamma$) 
and the doubly occupied ($n\sim 2$, $-\varepsilon_0>U+\Gamma$) 
regimes. Further details on the method 
used to calculate $G^R(\varepsilon+{\rm i}\eta)$ can be found in  
Refs. \cite{pruschke,gerace}. 

\begin{figure}[!ht]
\begin{center}
\includegraphics[width=\textwidth]{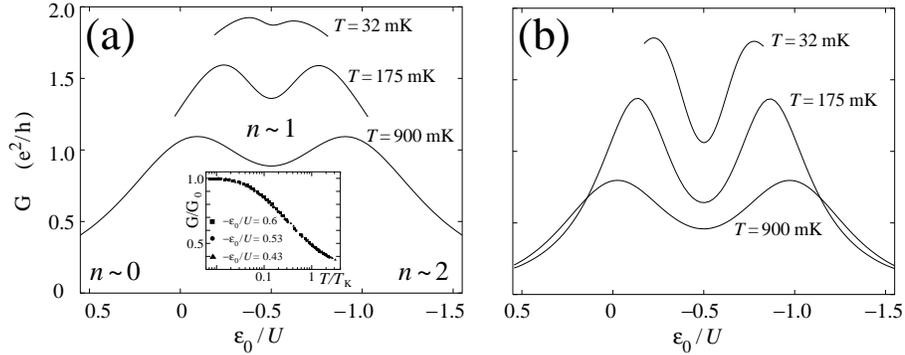}
\end{center}
\caption{Linear-response conductance
         as a function of $\varepsilon_0/U$ and for different 
         temperatures \cite{nota}.  
         (a) Parameters:
         $\Gamma=0.262$ meV and $U=0.8$ meV. 
         Inset: G$/$G$_0$ as
         a function of $T/T_K$ and for different choices
         of $\varepsilon_0/U$. Here
         G$_0$ is the value of G at the lowest temperature  
         that can be reached with the UNCA;
         $T_K$ is such that 
         G$(T_K)\equiv$G$_0/2$.
         (b) Parameters: 
         $\Gamma=0.120$ meV and $U=0.7$ meV.}
\label{fig1}
\end{figure}

The main purpose of the present work is
to understand the experimental data of Ref. \cite{silvano}.
The results of our (new) calculations 
are shown in Fig. \ref{fig1}ab for two 
different choices of parameters.
The parameters of Fig. 1a are our best fit
of the experimental data.
We notice that the conductance as a function of $\varepsilon_0/U$ is 
directly related to experimental data because of 
the linear relation between $\varepsilon_0$ and $V_g$.
The results in Fig. \ref{fig1}ab 
are in good agreement with previously reported 
exact 
numerical renormalization group (NRG) 
calculations (see Fig. 2 of Ref. \cite{izumida}).
The figure shows that, at high temperature
($T\gg T_K$, see e.g. $T=900$ mK),
there are two Coulomb blockade peaks at energies
$\varepsilon_0/U\sim 0$ and $\varepsilon_0/U\sim -1$, as observed 
experimentally by different groups \cite {silvano,sper,vanderwiel}. 
Each peak corresponds  
to the addition of one electron to those confined in the QD. 
When the temperature is lowered, the peaks
approach each other, and the conductance gradually increases
for $n\sim 1$, while it decreases for $n=0$ and $n=2$.
At very low temperature ($T\ll T_K$, see e.g. Fig. 1a, $T=32$ mK) 
the Coulomb blockade peaks merge into a plateau at G$\sim2e^2/h$, 
which is located in the parameter region for which $n\sim 1$.

In the Kondo regime, the conductance is expected to be
a universal function of $T/T_K$.  
In the inset of Fig. \ref{fig1}a we show the 
calculated universal curve G$/$G$_0$ for three 
different choices of $\varepsilon_0/U$
in the Kondo regime (that is $\varepsilon_0/U$ between
$-0.4$ and $-0.6$).
The ratio G$/$G$_0$ is proportional to ln$(T/T_K)$
for $T\sim T_K$. At very low temperature ($T\ll T_K$) 
G$/$G$_0\propto (T/T_K)^2$, as 
expected in the  Fermi-liquid regime \cite{hewson} and as found 
experimentally \cite{silvano,vanderwiel}.  

The conductance of the spin 1/2 Anderson model 
was calculated with the NRG technique in Ref. \cite{silvano},
and directly compared with the reported experimental data.
Although very good agreement was reached at low temperature,
it was shown that at high temperature the theoretical conductance 
strongly underestimates the experimental value.
For the same choice of parameters of the Anderson model, we find
the same discrepancy (see Fig. 1b and Fig. 2 of Ref. \cite{silvano}).
It was then suggested in Ref. \cite{silvano} 
that multilevel effects could be responsible of this discrepancy. 
Here we explored another possibility. We suggest
that the experimental data {\it can} be
fitted within the spin 1/2 Anderson model, 
provided that an appropriate choice of 
parameters is made.
We found the best agreement for $\Gamma=0.262$ meV 
(about two times larger than the one used in Fig. 1b)
and $U=0.8$ meV.
The results obtained with these optimal parameters 
are shown in Fig. 1a, 
and they are in excellent agreement with experiments
both at high and low temperature.

In conclusion, we have calculated the
conductance of a system made of a QD coupled
to two leads, described by the spin 1/2 Anderson model.
We adopted the UNCA, and we found good agreement with exact 
NRG results \cite{izumida} and experiments \cite{sper,vanderwiel}.
We have shown that recent experiments \cite{silvano}, 
contrarily to what previously proposed,
{\it can} be understood in the framework of the  
spin 1/2 Anderson model, in which the effects of the multi-level
structure of the dot is neglected.
Although the spin 1/2 Anderson model can be solved exactly 
by using the NRG, we believe that the UNCA is more suitable 
than NRG for extensions to realistic systems, and it
could become an important tool to interpret experiments
in which the realistic electronic structure of the 
QDs plays a crucial role (e.g. the Kondo effect in quantum dots
for integer spin \cite{sper2}).


\begin{thebibliography}{99}
\bibitem{nota}
At low temperature, in the empty orbital 
and in the doubly occupied regimes,
a spurious peak appears in the UNCA local DOS
close to the Fermi level. This is a known shortcoming
of the UNCA \cite{pruschke}.
Thus results at temperatures $T=175$ mK (Fig. 1a) 
and $T=32$ mK (Fig. 1ab) 
are shown only in the regions in which the UNCA is reliable.
\end{thebibliography}
\end{document}